\documentclass[runningheads]{llncs}
\usepackage{graphicx}
\usepackage{float}
\begin{document}
\title{Data Aggregation In The Astroparticle Physics Distributed Data Storage\thanks{Supported by the Russian Science Foundation, grant \#18-41-06003.}}
%
%
\author{
Minh-Duc Nguyen\inst{1}\orcidID{0000-0002-5003-3623}
\and Alexander Kryukov\inst{1}
\and Julia Dubenskaya\inst{1}
\and Elena Korosteleva\inst{1}
\and Igor Bychkov\inst{2}
\and Andrey Mikhailov\inst{2}
\and Alexey Shigarov\inst{2}
}
\authorrunning{M.D. Nguyen et al.}
%
\institute{Lomonosov Moscow State University, Skobeltsyn Institute of Nuclear Physics, 1/2 Leninskie Gory, 119991, Moscow, Russia \email{nguyendmitri@gmail.com}
\and Matrosov Institute for System Dynamics and Control Theory, Siberian Branch of Russian Academy of Sciences, Lermontov st. 134, Irkutsk, Russia
\email{shigarov@icc.ru}\\
%
}
\maketitle              
\begin{abstract}
German-Russian Astroparticle Data Life Cycle Initiative is an international project whose aim is to develop a distributed data storage system that aggregates data from the storage systems of different astroparticle experiments. The prototype of such a system, which is called the Astroparticle Physics Distributed Data Storage (APPDS), has been being developed. In this paper, the Data Aggregation Service, one of the core services of APDDS, is presented. The Data Aggregation Service connects all distributed services of APPDS together to find the necessary data and deliver them to users on demand.
\end{abstract}
\section{Introduction}
The amount of data being generated by various astroparticle experiments such as TAIGA \cite{bib_taiga}, KASCADE-Grande \cite{bib_kascade_grande}, MAGIC \cite{bib_magic}, CTA \cite{bib_cta}, VERITAS \cite{bib_veritas}, and HESS \cite{bib_hess} is tremendous. Processing data and, more importantly, delivering data from different experiments to end-users are one of the real issues in open science and particularly in open access to data. To address this issue, the German-Russian Astroparticle Data Life Cycle Initiative had been started in 2018. The primary goal of the project is to develop a prototype of a distributed storage system where data of two physical experiments, TAIGA and KASCADE-Grande, are aggregated in one place and to provide a unified access mechanism to end-users.

Such a prototype has been being developed and is called Astroparticle Physics Distributed Date Storage (APPDS). The design of the system is based on two key principles. The first one is creating no interference with existing data storage systems of the physical experiments. This principle is critical since most existing collaborations in astroparticle physics, not only the collaborations of TAIGA and KASCADE-Grande experiments, have been historically using their own stack of technologies and approach to data storage. Any change to the existing data storage systems might cause potential problems which are hard to be discovered. The second principle is using no computing resources at the sites of the storage systems to handle users' queries. This principle leads to creating a global metadata database where the data description from all storage systems is aggregated in one place. All searching and filtering operations are performed within the global metadata database. The operation results are delivered to users via a web interface. Actual data transfer from the existing data storage systems takes place only when users want to access the inside content of a file. Thus, APPDS causes no load to the computing resources at each site; the only load on the data storage systems is data delivery.

This article is organized as follows. In the second section, the architecture overview of APPDS is presented. Section 3 is dedicated to the design of the Data Aggregation Service, which is the core of APPDS, and the stack of technologies that have been used. In conclusion, the current state of the service is presented.

\section{APPDS architecture}

\begin{figure}
\includegraphics[width=\textwidth]{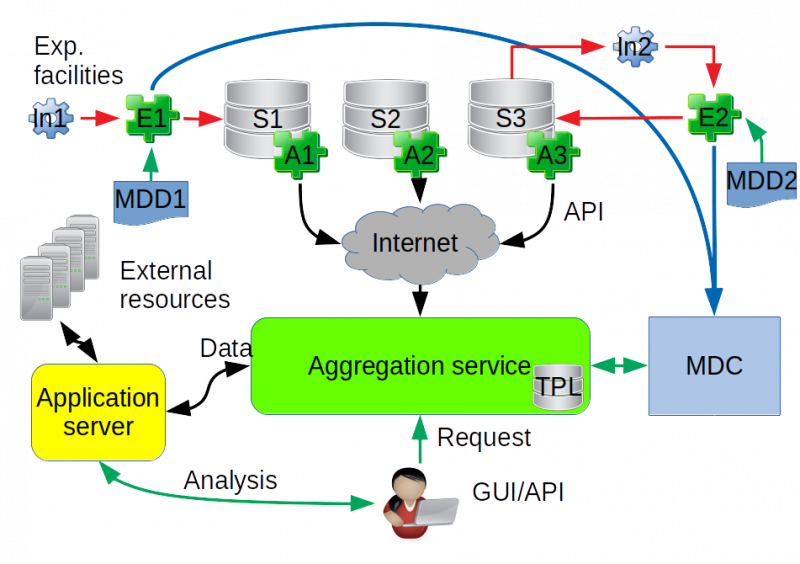}
\caption{APPDS architecture} \label{figure_appds_architecture}
\end{figure}

The architecture overview of APDDS is presented in Fig.\ref{figure_appds_architecture}. S1, S2, S3 are data storage instances of the physical experiments. In1 is the original data input. In2 indicates the case when the original data from In1, which are already stored in a storage instance, are being reprocessed. At the level of data storage instances, to preserve the original data processing pipeline, a program called Extractor (E1) is injected into the pipeline. In most cases, input data are files. After standard processing, the files are passed to the Extractor. The Extractor retrieves metadata from the files using the metadata description (MDD) provided by the development groups of the physical experiments, sends the metadata to the Metadata Database using its API, and passes the files back to the pipeline. If the data need to be reprocessed, the same pipeline is applied but with a different type of the Extractor (E2).

The files of each storage instance are delivered to the Data Aggregation Service by the Adapter, which is a wrapper of the CernVM-FS server \cite{bib_cernvmfs}.

To retrieve necessary files, the user forms a query using the web interface provided by the Data Aggregation Service. When the Data Aggregation Service receives the user query, it asks the Metadata Database for the answer. When the Metadata Database answers, the Data Aggregation Service generate a corresponding response and delivers it to the user.

All components of APPDS are talking to each other via RESTful API \cite{bib_restful_api}. 3rd party application services can also talk to APPDS via RESTful API. The key business logic of APPDS is implemented in the Data Aggregation Service whose design and implementation are considered in the next section.

\section {Data Aggregation Service}

\begin{figure}[h]
\includegraphics[width=\textwidth]{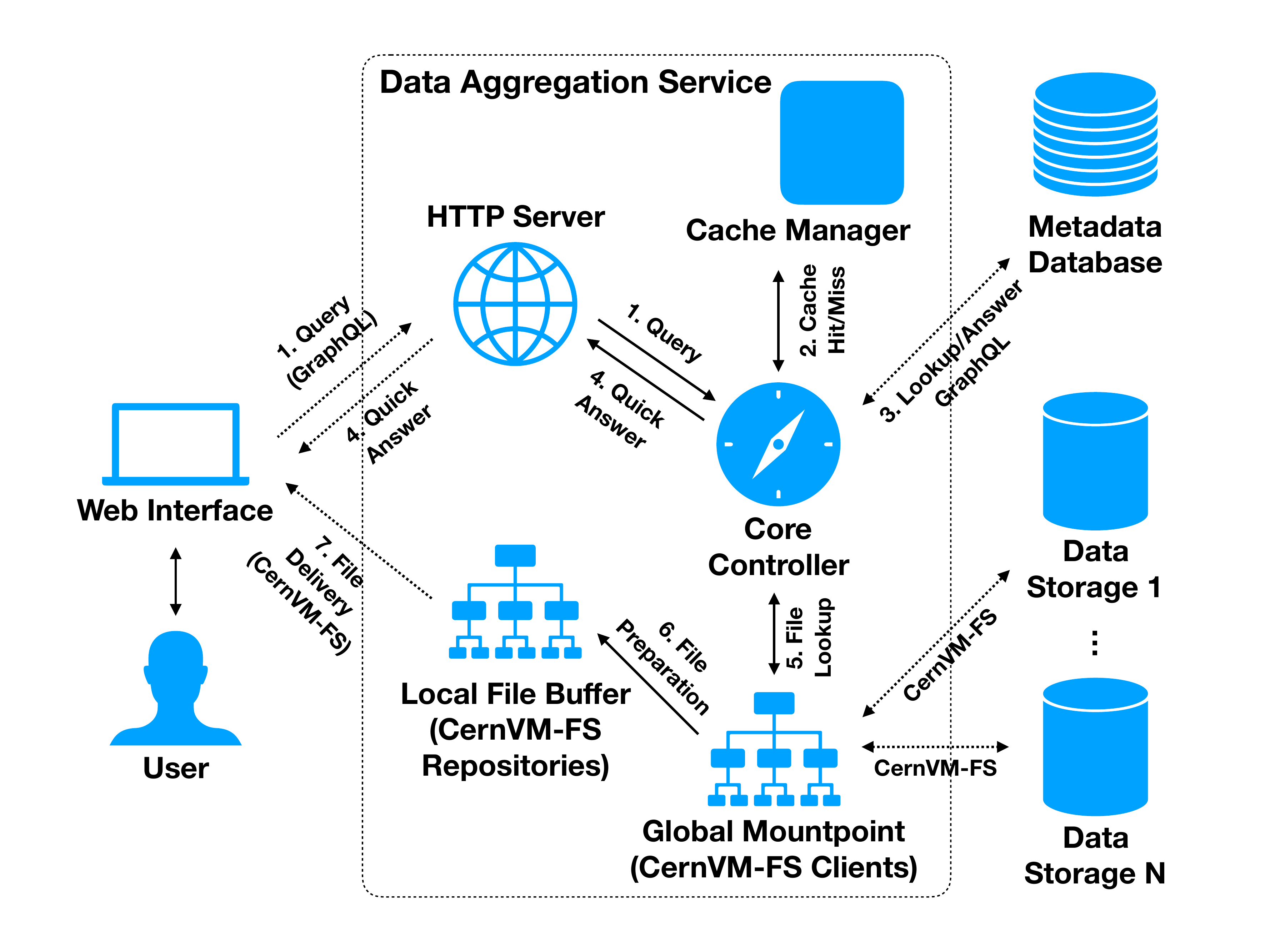}
\caption{Data Aggregation Service Design} \label{figure_data_aggregation_service_architecture}
\end{figure}

Figure \ref{figure_data_aggregation_service_architecture} shows how the components of the Data Aggregation Service are connected and how they interact with other services of APPDS. The user interacts with APPDS via the Web Interface provided by the Data Aggregation Service. Using the Web Interface, the user can make a query to search for necessary data accumulated by APPDS from all data storage instances. The key component of the Data Aggregation Service is the Core Controller. Whenever it receives a query, the following query processing pipeline is applied:

\begin{itemize}
    \item Query registration
    \item Cache lookup
    \item Metadata Database lookup
    \item Quick response to the user
    \item Full response generation and preparation for delivery
\end{itemize}

\subsection{Query Types}

There are two types of queries that a user can make using the Web Interface of Data Aggregation Service. Identification queries are used to retrieve the basic information about the resources provided by APPDS such as facilities, clusters, detectors, data channels, data authorities, permissions, and available data. Search queries are used when the user wants to look for the available data using a list of filters like data availability interval, energy range, facility location, detector types, data channel specification, weather condition, etc. Typical data lookup scenarios that users might create are:
\begin{itemize}
    \item data obtained by one facility or all available facilities for a certain period;
    \item season data which start from September to the end of May;
    \item data obtained in a testing period or a specific run;
    \item regular monthly or weekly data.
\end{itemize}

The obvious choice to implement the dialogue between the Web Interface and the Aggregation Service as separate microservices is to create a RESTful API. While identification queries are the best candidates to be implemented using a set of endpoints as in the standard RESTful API, the search queries with their complex filters are not. Using the standard RESTful API to implement the search queries leads to complex query sets containing redundant data that users do not need. The best approach to implement search queries is to use the Facebook GraphQL \cite{bib_graphql} instead of the standard RESTful API. By using GraphQL all typical search queries can be formatted flexibly as a JSON object that later is sent to the Metadata Database, and the responses to the queries contain the exact information that the user needs with no redundancy. It is also quite easy to implement additional filters without breaking compatibility using GraphQL.

\subsection{Query Processing Pipeline}

Whenever the Core Controller receives a query from the GraphQL Backend, it calculates the query checksum using the MD5 algorithm. The checksum is used as the query ID. After that, the Core Controller check against the Cache Manager to find if such a query is already registered. If not the Core Controller registers the query with the Cache Manager. The Cache Manager is implemented based on Redis \cite{bib_redis}, a popular caching mechanism. The query body is converted into a Redis hash which is a map composed of fields corresponding to the filters, each of which contains the filter value. 

If the response to the query is already cached and the active period does not expire, it will be delivered to the user. If the response to the query has not been cached, the Core Controller forwards the query to the Metadata Database. The Metadata Database generates a SQL-query according to the original query and executes it. Depending on the data sources, the response from the Metadata Database can vary. The response, including data from the TAIGA experiment, is a set of files which match the filters specified in the original query. The Metadata Database also calculates the checksum of the response, adds it to the final response and sends it to the Core Controller. The Core Controller, in turn, appends the response to the query in the Cache Manager. After that, the Core Controllers sends the response to the Web Interface so that the user could look at it quickly.

At the same time, the Core Controller starts preparing the full response for delivery. Files from all Data Storage instances are exported to the Data Aggregation Service as separate CernVM-FS repositories. All repositories are located inside a Global Mount Point where each subdirectory is the mount point of each repository. To prepare the data, the Core Controller finds the files in the Global Mount Point and copies them to a directory in the Local File Buffer. The directory is then configured as a CernVM-FS repository for export. The name of the repository is the same as the query ID. If the user query requires not the original files but a composition of them, the Core Controller scans through the original files, composes a new set of files, and puts them into the directory. Later, if the user wants to work with the files locally, she can mount the prepared CernVM-FS repository to her computer. The Core Controller also generates an archive containing all files. The archive is put into the repository and can be downloaded separately. If the full response to a query is not used more than a specific time, it will be deleted from the Local File Buffer and the Cache Manager.

\section{Conclusion}

Currently, a beta version of APPDS, including the Data Aggregation service, the Metadata Database, the Adapter, and the Web Interface, has been implemented and being tested against the storage system of the TAIGA experiment. In the next release, support for the storage system of the KASCADE-Grande experiment will be included. The first working prototype of the system is planned to be released this fall. The proof-of-concept tests and performance benchmarks are also planned after the release.


\begin{thebibliography}{8}
\bibitem{bib_taiga}
Bundev N. et al.: The TAIGA experiment: From cosmic-ray to gamma-ray astronomy in the Tunka valley. In: Nuclear Instruments and Methods in Physics Research Section A: Accelerators, Spectrometers, Detectors and Associated Equipment February 2017, vol. 845, pp 330-333. \doi{10.1016/j.nima.2016.06.041}

\bibitem{bib_kascade_grande}
Apel W.D. et al.: The KASCADE-Grande Experiment. In: Nuclear Instruments and Methods in Physics Research Section A 620 April 2010: pp 202-216. \doi{10.1016/j.nima.2010.03.147}

\bibitem{bib_magic}
Anderhub H. et al.: MAGIC Collaboration. In: 31st International Cosmic Ray Conference (ICRC 2009). \doi{10.15161/oar.it/1446204371.89}

\bibitem{bib_cta}
Cherenkov Telescope Array: Exploring the Universe at the Highest Energies.
\url{https://www.cta-observatory.org/}

\bibitem{bib_veritas}
Very Energetic Radiation Imaging Telescope Array System
\url{https://veritas.sao.arizona.edu}

\bibitem{bib_hess}
High Energy Stereoscopic System. \url{https://www.mpi-hd.mpg.de/hfm/HESS/}

\bibitem{bib_cernvmfs}
Buncic P., Aguado Sanchez C., Blomer., Franco L., Harutyunian A., Mato P., Yao Y.: CernVM – a virtual software appliance for LHC applications. In: Journal of Physics: Conference Series 219 (2010) 042003. \doi{10.1088/1742-6596/219/4/042003}

\bibitem{bib_restful_api}
Fielding R. Th.: Architectural Styles and the Design of Network-based Software Architectures. \url{https://www.ics.uci.edu/~fielding/pubs/dissertation/fielding\_dissertation.pdf}

\bibitem{bib_graphql}
Facebook Inc.: GraphQL. \url{https://graphql.org}

\bibitem{bib_redis}
Redis Labs: Redis documentation. \url{https://redis.io/documentation}

\end{thebibliography}
\end{document}